\documentclass[aps,prd,amsmath,nofootinbib,floatfix,fleqn]{revtex4}
\pdfoutput = 1

\usepackage{color}
\usepackage[dvipsnames]{xcolor}
\usepackage{graphicx}
\usepackage{bm}
\usepackage[utf8]{inputenc}
\usepackage{subfig}
 \usepackage{amssymb}
 \usepackage{multirow}
 
 \usepackage{float}
 \usepackage{epsfig}% need for subequations
\usepackage{pdfpages}

\definecolor{darkblue}{rgb}{0.0,0,0.5}
\definecolor{darkgreen}{rgb}{0.0,0.3,0.0}
\definecolor{redish}{rgb}{0.675,0,0.2}
\definecolor{red}{rgb}{0.8,0,0}
\definecolor{green}{rgb}{0,0.6,0}
\definecolor{blue}{rgb}{0,0,0.8}

\usepackage[unicode=true, bookmarks=false, linkcolor = darkblue, citecolor = redish, breaklinks=false, colorlinks=true, hyperfootnotes=true]{hyperref}

\newcommand{\open}{\sphericalangle}

\begin{document}

\date{\today}
%\title{Insights into the higher-twist distribution $e(x)$ at CLAS}
\title{Extraction of the higher-twist parton distribution $e(x)$ from CLAS data}

\preprint{}

\author{Aurore Courtoy$^{(a,1)}$ and Angel Miramontes$^{(b,2)}$, Harut Avakian$^{(c)}$, Marco Mirazita$^{(d)}$, Silvia Pisano$^{(d, e)}$}
\email{$^{1}$aurorecourtoy@gmail.com, $^{2}$angel-aml@hotmail.com}

\affiliation{$^{(a)}$Instituto de F\'isica, Universidad Nacional Aut\'onoma de M\'exico, Apartado Postal 20-364, 01000 Ciudad de M\'exico, M\'exico \\
$^{(b)}$ Instituto de F\'isica y Matem\'aticas, Universidad Michoacana de San Nicol\'as de Hidalgo, Morelia, Michoac\'an 58040, Mexico\\
$^{(c)}$ Thomas Jefferson National Accelerator Facility, Newport News, Virginia 23606\\
$^{(d)}$ INFN, Laboratori Nazionali di Frascati, 00044 Frascati, Italy\\
$^{(e)}$ Centro Fermi - Museo Storico della Fisica e Centro Studi e Ricerche “Enrico Fermi", Rome, Italy}

 \date{\today}

\begin{abstract}
This manuscript presents the new analysis of recent CLAS and CLAS12 data aimed at the extraction of $e(x)$. This publication benefits from a proof-of-principle analysis of a smaller preliminary data set published in a preprint form in \url{1405.7659}, \cite{Courtoy:2014ixa}.
\\
    
    We present the first point-by-point extraction of a twist-3 PDF. The scalar PDF, $e(x)$, is accessed through the  analysis of the data for the $\sin\phi_R$-moment of the beam-spin asymmetry for  dihadron production in semi-inclusive DIS off proton target at CLAS and CLAS12. The dihadron formalism allows for use of collinear framework, hence calling for a minimal set of approximations and hypotheses. The  extracted  PDF $e(x)$ carries insights into the physics of the largely-unexplored quark-gluon correlations, 
    and its first Mellin moment is related to the marginally-known scalar charge of the nucleon. We show that the proton flavor combination of the scalar PDF is non-zero at more than $74\%$ probability.
\end{abstract}

\maketitle

%%%%%%%%%%%%%%%%%%%%%%%%%%%%%%%%%%%%%%%%%%%
\section{Introduction}
\label{sec:intro}

The recent advent of mid-energy experiments, relevant to explore the structure of hadrons, has led to an increased focus on the dynamics of QCD. It is at mid-values of the scale probing hadrons, $Q$, at the limit between the perturbative and the non-perturbative regimes,  that non-perturbative contributions play a decisive role. As the observation of unexpected large transverse polarization effects opened the door for studying transverse momentum dependence of distribution functions, experiments designed access to semi-inclusive processes. Those shed light on unexplored collinear PDFs~\cite{Artru:1989zv} at the same time they proposed insights into the 3D structure of hadrons~\cite{Ralston:1979ys}. But the kinematics of semi-inclusive DIS also made collinear subleading terms relevant. As such, those were thought as natural candidates to clarify the origin of large transverse polarization effects in observables, an hypothesis that is still explored today, {\it e.g.}~\cite{Avakian:2019drf,Cammarota:2020qcw}. Higher-twist corrections are understood in phenomenology as terms that are suppressed {\it w.r.t.} the dominant contribution. While those are expressed as terms of the order ${\cal O}(M/Q)$, with $M$ the target mass and $Q$ the hard scale, there exist various extensions of the term {\it higher twist}, including  the definitions of the field theoretical objects or the overall suppression in an observable. 

Nowadays, the relation between particular TMD PDFs and twist-3 collinear PDFs is understood from the expression of extra degrees of freedom in both approaches. Higher-twist PDFs enclose non-perturbative information related to multiparton correlations. The structure of the matrix elements that define PDFs reveals the various contributions to the underlying physical picture. In a first time, twist-3 PDF can be reduced to an expression of the leading-twist PDF with whom they share the Dirac structure~\cite{Wandzura:1977qf}. Departure from that approximation were suggested already in early evaluations in models for hadron structure, and then through phenomenology~\cite{Accardi:2009au}. 
% 22-03-21

The twist-3 parton distribution function $e^q(x)$ is of general interest since it encapsulates key information on quark-gluon-quark correlations in the nucleon. Its contribution to the scalar quark-quark bilocal operator has been highlighted through the study of QCD equations of motion, {\it e.g.}~\cite{Kodaira:1998jn} -- see Ref.~\cite{Efremov:2002qh} for a review. %Additionally, t
The distribution $e(x)$, through the nucleon-sigma terms, plays an important role in the understanding of the decomposition of the nucleon mass in terms of contributions from gluons and quarks~\cite{Ji:2020baz}, for which different mass decomposition schemes from the energy-momentum tensor exist in the literature -- {\it e.g.}~\cite{Ji:1994av,Ji:2021mtz,Lorce:2017xzd,Lorce:2021xku}. 
That term originates from the singularity induced by zero modes in the light-cone formalism~\cite{Burkardt:2001iy,Aslan:2018tff,Ji:2020baz,Hatta:2020iin,Bhattacharya:2020jfj}.
The scalar PDF has been studied in non-perturbative models for hadron structure~\cite{Jaffe:1991ra,Wakamatsu:2000fd, Wakamatsu:2003uu, Schweitzer:2003uy,Mukherjee:2009uy,Avakian:2010br,Lorce:2014hxa,Pasquini:2018oyz,Bastami:2020rxn}. In most quark models, the dominant contribution is found to be the mass contribution, which originates from the equation of motions for free fields, {\it e.g.}~\cite{Lorce:2014hxa}. The three contributions to the scalar PDF are outlined as follows
\begin{eqnarray}
e^q(x)&=&e_{\mbox{\tiny{sing.}}}^q(x)+e_{qgq}^q(x)+e_{\mbox{\tiny{mass}}}^q(x)%\,. \label{eq:def_ex}
\end{eqnarray}

The contribution from the sigma terms to the proton mass, and hence answers on its decomposition, is one of the research problems to be investigated in the Electron-Ion Collider~\cite{AbdulKhalek:2021gbh}. 
This goes hand-in-hand with parallel efforts to understand the emergence of hadronic mass  through the breaking of chiral symmetry in the pion sector~\cite{Adams:2018pwt, Aguilar:2019teb}.

Its chirally-odd nature makes the distribution $e(x)$ challenging to access in experiments, as it can contribute to an observable only in connection with another chirally-odd distribution. A first extraction in the TMD framework was performed in Ref.~\cite{Efremov:2002ut}. In Ref.~\cite{Bacchetta:2003vn}, it was shown that $e(x)$ can be extracted, in a collinear framework, through spin asymmetries in semi-inclusive DIS which has been measured in CLAS collaboration.

We here propose the first  phenomenological analysis of the scalar PDF in a collinear framework, using both CLAS and CLAS12 data for dihadron production in semi-inclusive DIS off (unpolarized) proton target~\cite{CLAS:2020igs,Hayward:2021psm}. The  point-by-point extraction is made possible through the knowledge of dihadron fragmentation functions (DiFFs) from Belle data~\cite{Bacchetta:2011ip,Courtoy:2012ry,Radici:2015mwa}. The full analysis required complementary data on longitudinally-polarized targets from COMPASS~\cite{Sirtl:2016,Sirtl:2017rhi}.

The manuscript is organized as follows. In Section~\ref{sec:formalism}, we describe the dihadron formalism for the beam-spin asymmetry and define all relevant players. The core of the analysis is presented in Section~\ref{sec:extraction}. Reconstruction of the dihadron-relevant projections of the asymmetry provides for a useful yet short benchmarking of our analysis. It is followed by the extraction {\it per se}. That section is supplemented by two extensive appendices, App.~\ref{sec:BSA} and~\ref{sec:COMPASS}. The interpretation of the results is discussed in Section~\ref{sec:disc}. Finally, we draw conclusions and discuss forthcoming extensions of this work in Section~\ref{sec:concl}.

%%%%%%%%%%%%%%%%%%%%%%%%%%%%%%%%%%%%%%%%%%%
\section{Formalism}
\label{sec:formalism}

In this analysis, we consider the structure function $F_{LU}^{\sin\phi_R}$ given in Appendix~\ref{sec:BSA}, corresponding to an unpolarized target,  for $\pi^+\pi^-$ pair production in semi-inclusive DIS.  
For the longitudinal polarization of the beam,  the relevant azimuthal asymmetry is 
\begin{eqnarray} %\left
A_{LU}^{\sin \phi_R } \left( z, M_h, x ; Q, y \right)
&=&
\frac{\frac{4}{\pi}\sqrt{2\,\varepsilon (1-\varepsilon)} \int d \cos \theta \, F_{LU}^{\sin\phi_R}}
{\int d \cos \theta\, \left( F_{UU,T}+ \epsilon F_{UU,L} \right) }%\quad,
\label{e:ssamaster}
\end{eqnarray} 
where $\varepsilon$ is the ratio of longitudinal and transverse photon flux and can be expressed in terms of $y$. 
To leading-order in $\alpha_s$ and leading term in the partial wave expansion (see App.~\ref{app:diff}), the beam-spin asymmetry (BSA)  becomes~\cite{Bacchetta:2003vn} %
\begin{eqnarray}
&&A_{LU}^{\sin \phi_R } \left( x,  z, M_h ; Q, y \right) \nonumber\\
&=&-\frac{W(y)}{A(y)}\,\frac{M}{Q}\,\frac{|\bf{R}  |}{M_h} \, 
\frac{ \sum_q\, e_q^2\, \left[ x e^q(x, Q^2)\, H_{1, sp}^{\sphericalangle, q}(z, M_h, Q^2)  + \frac{M_h}{z M} \,  f_1^q(x, Q^2)\, 
          \tilde{G}_{sp}^{\sphericalangle, q}(z, M_h, Q^2) \right]  } 
       { \sum_q\, e_q^2\,f_1^q(x, Q^2)\, D_{1,ss+pp}^q (z, M_h, Q^2) }%\quad ,
       \nonumber\\
 \label{eq:alu}
\end{eqnarray}
with $A(y)$, $W(y)$ and $B(y)$ (in App.~\ref{app:diff}) the depolarization factors. %We drop the dependence in $y$ on the {\it l.h.s.} of the asymmetries as

The dependence in $(z, M_h)$ is factorized in the DiFFs and kinematical factors, leaving the dependence in $x$ for the PDFs.
The twist-2 functions are $f_1(x), H_1^{\open} (z, M_h)$ and $D_1(z, M_h)$, while the twist-3 functions are $e(x)$ and $\tilde{G}^{\sphericalangle}(z, M_h)$. 
From now on, we will drop the indices referring to the partial waves.
\\

The CLAS Collaboration recently  collected data on BSA for dihadron production in semi-inclusive DIS by impinging the CEBAF $5.5$-GeV longitudinally-polarized
electron beam on an unpolarized $^2H$ hydrogen target~\cite{CLAS:2020igs} as well as with the CLAS12 spectrometer using a 10.6 GeV longitudinally spin-polarized electron beam~\cite{Hayward:2021psm}.
The CLAS data offer a unique access to the scalar PDF, $e(x)$, by means of an analysis of the fragmentation functions at leading and subleading twist.  Dihadron fragmentation functions  have been studied at Belle. A fit of the $e^+e^-\to (\pi^+\pi^-) (\pi^+\pi^-)X$ multiplicities fixed the unpolarized DiFFs\footnote{Data for those multiplicities are now available in Ref.~\cite{Belle:2017rwm}, but require a complete new analysis, possibly at NLO.}~\cite{Courtoy:2012ry}.
The chiral-odd DiFFs were hence analyzed from the Artur-Collins asymmetry at Belle~\cite{Belle:2011cur}, using the Hessian representation of uncertainties~\cite{Bacchetta:2012ty} -- used in a preliminary version of this work -- and the bootstrap method~\cite{Radici:2015mwa}. In this paper, we will use the latter set for the chiral-odd DiFFs. Definitions and details about the relevant DiFFs are given in App.~\ref{app:diff}.

The point-by-point extraction of the scalar PDF would be technically analogous to the first collinear extraction of the transversity PDF~\cite{Bacchetta:2011ip} were it not for the second term on the {\it r.h.s.} of Eq.~(\ref{eq:alu}).
Our analysis of the twist-$3$ parton distribution relies on the treatment of that second term, that is a multiparton dependence from the fragmentation part.
Up to date, there is no phenomenological study of twist-$3$ dihadron fragmentation functions. 
Only a few model evaluations have been performed recently~\cite{Luo:2019frz,Yang:2019aan}, mainly extending the pioneering evaluation of DiFFs in the spectator model~\cite{Bacchetta:2006tn}.
 To bypass the absence of rigorous phenomenological fragmentation functions at higher-twist,  we will consider two conceptually different scenarios to make the most of the information at hand. That information consists of the two CLAS data sets, the preliminary COMPASS data on longitudinally-polarized target for semi-inclusive dihadron production~\cite{Sirtl:2016, Sirtl:2017rhi} as well as guidance from the model evaluations mentioned above.

To this aim, we will use all three one-dimensional projections of the data on the $(x,z, M_h)$ variables. Those call for --truncated-- integrals over the kinematical range of the experiment, {\it i.e.}
\begin{eqnarray}
\nonumber\\
n_{q}^{\mbox{\scriptsize FF}}(Q^2) &=& \int_{z_{\text{\tiny min}}}^{z_{\text{\tiny max}}} dz \,\int_{m_{\pi\pi, \, \text{\tiny min}}}^{m_{\pi\pi, \, \text{\tiny max}}}
 dM_h \, \mbox{FF}^q (z, M_h; Q^2)  %\; , 
 \label{e:nq} \\
 n_{q,\, m}^{\mbox{\scriptsize PDF}}(Q^2) &=& \int_{x_{\text{\tiny min}}}^{x_{\text{\tiny max}}} dx \, x^{m}\,\mbox{PDF}^q(x;Q^2)%\,,
 \end{eqnarray}
 where $\mbox{FF}$ refers to $D_1$ for $n_{q}(Q^2) $, $|{\bf R}|/M_h\times H_{1}^{\open}$ for $n_{q}^{\uparrow}(Q^2) $ and $|{\bf R}|/M\times {\tilde G}^{\open}$ for $n_{q}^{\tilde{G}^{\sphericalangle}}(Q^2) $. The $x$-dependence of the asymmetry comes solely from the PDF through the $m^{\mbox{\scriptsize th}}$-truncated moment. 
 Those quantities are discretized to the values corresponding to each bin, with $i=1,\cdots n_{\mbox{\tiny set}}$ where $n_{\mbox{\tiny set}}$ corresponds to the number of data points in a given set. Specifically, the CLAS data is composed of three bins, and the CLAS12, 12 bins. For one-dimensional projections on a fragmentation function variable, we average over the bin boundaries for that given variable.\footnote{We have checked that the difference between using the averaged functions and the functions for the average kinematical value was well within the experimental uncertainties for the $z$ projections and analyzed the physical meaning of the average $M_h$ values with care.} 
 At CLAS, the $M_h$ binning has been chosen to  consistently include the relevant resonances from the two-pion production, such as the $\rho$ meson. Given the small values of the experimental $Q^2$ values, the analysis will be  carried out without QCD evolution for the PDFs and the FFs. All quantities are considered at $Q_0=1$ GeV. We therefore drop the scale dependence of these quantities.

The triptych of one-dimensional projections can be factorized into the PDF-relevant and FF-related projections. Of interest for this study is the $x$-projected asymmetry, for which Eq.~(\ref{eq:alu}) can be expressed as
\begin{eqnarray}
A_{LU}^{\sin \phi_R } \left( x_i ; Q_i, y_i \right)&=&-\frac{W(y_i)}{A(y_i)}\,\frac{M}{Q_i} \,\frac{  x_i  \left[ \frac{4}{9} e^{u_V}(x_i)- \frac{1}{9} e^{d_V}(x_i)\right] \,n_{u, i}^{\uparrow}  +   \left[\frac{4}{9} f_1^{u_V}(x_i)-\frac{1}{9} f_1^{d_V}(x_i)\right]/z_i\,
         n_{u, i}^{\tilde{G}^{\sphericalangle}}   } 
       {\sum_{q=u,d,s}\, e_q^2\,f_1^q(x_i , Q_i ^2)\, n_{q, \, i}  }%\quad.
       \nonumber\\
       \nonumber\\
       \nonumber\\
      & =&-\frac{W(y_i)}{A(y_i)}\,\frac{M}{Q_i}\frac{x_i e^{\rm P}(x_i)\, n_{u,\, i}^{\uparrow}  + f_1^{\rm P} (x_i)\, /z_i\;
         n_{u,\, i}^{\tilde{G}^{\sphericalangle}}  }{ \sum_{q=u,d,s}\, e_q^2\,f_1^q(x_i)\, n_{q, \, i}  }%\quad,
       \label{eq:alu_x_fl}
\end{eqnarray}
where we have used the relations~(\ref{eq:D1_CC}, \ref{eq:H1_IS}) and neglected the strange quark contributions in the denominator. We have defined  the proton-flavor combination $f^{P}=4/9\, f^{u_{\mbox{\tiny V}}}-1/9\, f^{d_{\mbox{\tiny V}}}$, with $f^{q_{\mbox{\tiny V}}}\equiv f^{q}-f^{\bar q}$.

We can now move on to analyze the beam-spin asymmetry within our two scenarios.
The first, called the {\it Wandzura-Wilzcek} scenario, uses the  reduction of twist-$3$ PDF to a twist-$2$ partner, disregarding the dynamical contributions encoded through the $qgq$ and mass terms as well as the singularities.  
In this scenario, only the first term on the {\it l.h.s.} of the BSA, Eq.~(\ref{eq:alu_x_fl}), contributes. The interpretation of the thus-obtained $e(x)$ results to be slightly awkward, as $e(x)$ vanishes in the Wandzura-Wilzcek approximation. This is why we shall prefer to call it the ``$0^{th}$ approximation," allowing then for a wider range of physical meaning. Account for dynamical contributions from the fragmentation sector is intended in a second scenario, beyond the  ``$0^{th}$ approximation," for which we need to infer the order of magnitude of the twist-3 sector of DiFFs. This is discussed in detail in the next section.

%%%%%%%%%%%%%%%%%%%%%%%%%%%%%%%%%%%%%%%%%%%
\section{Point-by-point extraction}
\label{sec:extraction}

Having defined all relevant quantities, we can now focus on the technicalities of the extraction.
To study the impact of the statistical error, we employed the so-called bootstrap method, already used in Ref.~\cite{Radici:2015mwa}. This method consists in generating $N_{\rm rep}$ replicas of the $n$ starting data sample. In each replica, the original data point is perturbated by a Gaussian noise. Then, each replica describes a possible outcome from an independent experimental measurement. In this case, the number of replicas is fixed to $N_{\rm rep} = 104$ by the available set of DiFF fits.
That is, at $Q_0=1$ GeV, we have $N_{\rm rep}=104$ parameter sets for the chiral-odd DiFF. In the same fashion as has been done in Refs.~\cite{Radici:2015mwa,Benel:2019mcq}, the uncertainties will be propagated by generating $N_{\rm rep}$ replicas of the asymmetry data through a normal distribution given by the quoted one-sigma uncertainties -- we add systematic and statistical uncertainties in quadrature.
Together with the uncertainty coming from the dihadron fragmentation functions, they constitute the main contribution from known objects to the error band on $e(x)$.  The impact of further sources of uncertainties will be discussed here below. In particular, the bias coming from the determination of the twist-3 DiFF will matter greatly. The inclusion of the strange to the unpolarized cross section, as well as uncertainty coming from the unpolarized PDFs will be minor. The former is not considered in this work. 

%%%%%%%%%%%%%%%%%%
\begin{table}[tb]
\renewcommand{\arraystretch}{2}
\begin{tabular}{ |p{3cm}|p{3cm}|p{3cm}|p{3cm}| } 
\hline
&$x$  & $z$ & $M_h$ [GeV] \\
\hline
CLAS    & $[0.114, 0.593]$  & $[0.530, 0.948]$& $[2m_{\pi}, 1.734]$
\\
\hline
\multirow{2}*{CLAS12}  
            & \multirow{2}*{$]0,1]$} 
                        &\multirow{2}*{$[0.304, 0.872]$}    & $[2m_{\pi}, 0.63]$     \\
 \cline{4-4} 
 &  &    &$[0.63, 2.5]$ \\
\hline
\end{tabular}
\caption{Kinematics for the three relevant variables, from CLAS~\cite{CLAS:2020igs} and CLAS12~\cite{Hayward:2021psm}.}
\label{tab:kin}
\end{table}

The values of the kinematical variables $(x, z, M_h)$ for both data sets are shown in Tab.~\ref{tab:kin}.  In  the experimental CLAS12 analysis~\cite{Hayward:2021psm}, two separate invariant-mass regions have been defined, originally to study the transverse momentum of the final state for transverse-momentum dependent sensitive observables. It is justified by the appearance of vector mesons for $M_h>0.63$ GeV whose  contribution to the dihadron fragmentation process is understood to dominate.

%%%%%%
\subsubsection{Reconstruction of the $(z, M_h)$ projections}
\label{subsec:zmpipi}

The factorization of the variables corresponding to the distribution functions, $x$, to that related to the fragmentation, $(z, M_h)$, is evident from Eq.~(\ref{eq:alu}). Dihadron-related observables are often presented as a triptych of projections, in which the PDFs explicitly show in the $x$-dependent projection and is simply a normalization factor in the other two projections. From those $(z, M_h)$ projections, the behavior of the invariant mass will be particularly instructive to identify the underlying physics.

\begin{figure}[tb]
\includegraphics[scale=0.55]{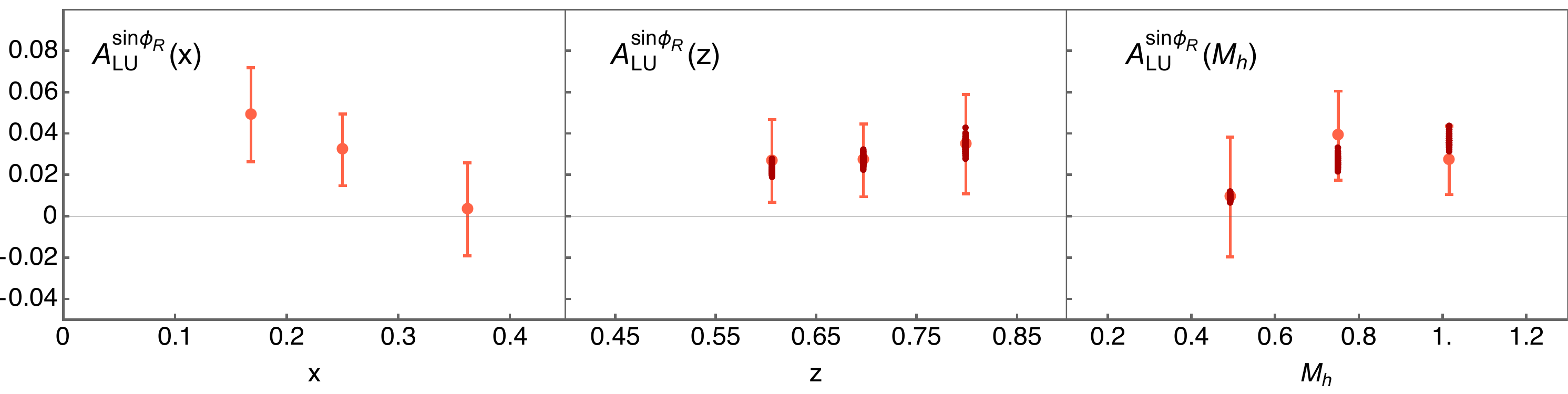}
\caption{The triptych for the asymmetry $A_{LU}^{\sin\phi_R}$ from CLAS~\cite{CLAS:2020igs} assuming leading-twist DiFFs~\cite{Radici:2015mwa}. Light-red error bars represent the CLAS data. %The cloud of red points correspond to the normally-generated replicas that will be used in the analysis. 
The burgundy points represent the reconstructed $z$ and $M_h$ projections, multiplied by a normalization factor for the $x$-integral (see text).}
\label{fig:zMhtriptychCLAS6}
\end{figure}

\begin{figure}[tb]
\includegraphics[scale=0.575]{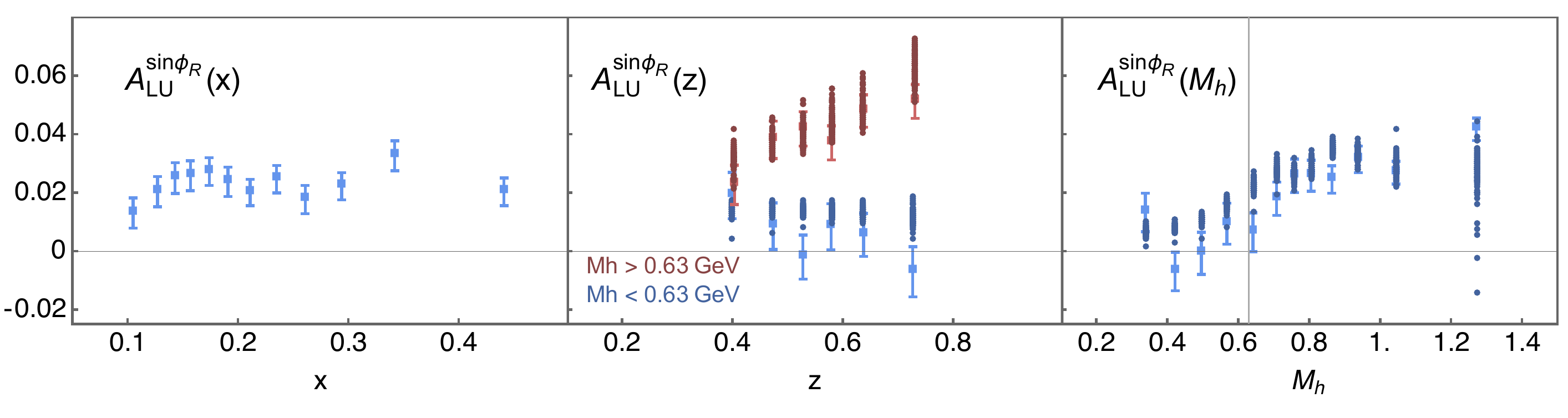}
\caption{The triptych for the asymmetry $A_{LU}^{\sin\phi_R}$ from CLAS12~\cite{Hayward:2021psm} assuming leading-twist DiFFs~\cite{Radici:2015mwa}. Light-blue error bars represent the CLAS12 data. The blue points represent the reconstructed $z$ and $M_h$ projections, multiplied by a normalization factor for the $x$-integral (see text). CLAS12 data are split into $M_h<$ or $>0.63$GeV, as shown in the $z$-projection.}
\label{fig:zMhtriptychCLAS12}
\end{figure}

As a first step on our analysis, we reconstruct the projections based on the analysis of leading-twist dihadron fragmentation from Belle~\cite{Radici:2015mwa}.
Using the $M_h$ and $z$ projections will serve as a check of the behavior of the fragmentation functions. The unknown $x$-contribution being integrated, it will contribute through a normalization factor, $n^{e}_{{\mbox{\tiny P}}, 2}$, that will be discussed in the next sections. In Figs.~\ref{fig:zMhtriptychCLAS6} and~\ref{fig:zMhtriptychCLAS12}, we show the BSA at CLAS and CLAS12, respectively, together with the  asymmetry  reconstructed from
\begin{eqnarray}
A_{LU, \mbox{\tiny reco.}}^{\sin \phi_R } \left(  M_{h, i},  z_i; Q_i, y_i \right)
&=&-\frac{W(y_i)}{A(y_i)}\,\frac{M}{Q_i}\,\frac{n^{e}_{{\mbox{\tiny P}}, 2}}{n^{f_1}_{{\mbox{\tiny P}}, 1}} \;  \, 
\frac{ n_{u, i}^{\uparrow}  } 
       { n_{u, i}  }
 \label{eq:mhz}
\end{eqnarray}
 for which the integrals $n_i$ correspond, for the central panels of Figs.~\ref{fig:zMhtriptychCLAS6} and~\ref{fig:zMhtriptychCLAS12}, to a projection onto $M_h$ and average in $z$ bins of the DiFF functional forms, or vice versa for the right panels. 
 %%%%%%%%%%%%%%%%%%%%%%%%%%%%%%%%%%%%%%%%%%
%%%%%%%%%%%%%%%%%%%%%%%%%%%%%%%%%%%%%%%%%%%%%%%%%%%%%%%%%%%%%%%%%%%%%%%%%%%%%%%%%%%
%%%%%%%%%%%%%%%%%%%%%%%%%%%%%%%%%%%%%%%%%%

 To get an estimate of the scaling factor $n^{e}_{{\mbox{\tiny P}}, 2}$,  we have minimized the chi squared between the data of the $M_h$ and $z$ projections and the BSA from Eq.~\eqref{eq:mhz}.  Using the data from CLAS and CLAS12, we obtain, respectively, %
 \begin{eqnarray}
 n_{{\mbox{\tiny P}}, 2}^{e,\, \mbox{\tiny CLAS}} &=& 0.056\pm 0.005\nonumber\\
 n_{{\mbox{\tiny P}}, 2}^{e,\, \mbox{\tiny CLAS12}} &=& 0.116\pm 0.042
 \label{eq:nx}
 \end{eqnarray}
 A difference between the two sets was expected given the coverage in $x$ of both data sets -- reported in Table~\ref{tab:kin}.
 
 Overall, the reconstruction from the DiFF fits are compatible, within error bars, with the experimental data. In particular, the behavior of the invariant mass projection suggests that, if there is a contribution from twist-3 DiFF to the beam-spin asymmetry, its trend must be either small and/or similar to that of $H_1^{\sphericalangle}$.
 We encountered difficulties in reproducing the asymmetries in $z$ for the lowest invariant-mass settings of CLAS12. The uncertainty on the $x$-integral is consequently increased.

%%%%%%
\subsubsection{$0^{th}$-approximation scenario}
\label{subsec:ww}

After this short benchmarking, we turn to the $x$ dependence of the asymmetry. We first work in the $0^{th}$-approximation scenario. This approximation originates from an early enthusiasm for a clean extraction within well-justified physical approximations to drop to twist-3 DiFF term altogether. %
 Then, Eq.~(\ref{eq:alu_x_fl}) consists in a single term and a point-by-point extraction results in the same fashion as what was proposed in Ref.~\cite{Bacchetta:2011ip}. 

\begin{figure}[tb]
\includegraphics[scale=1]{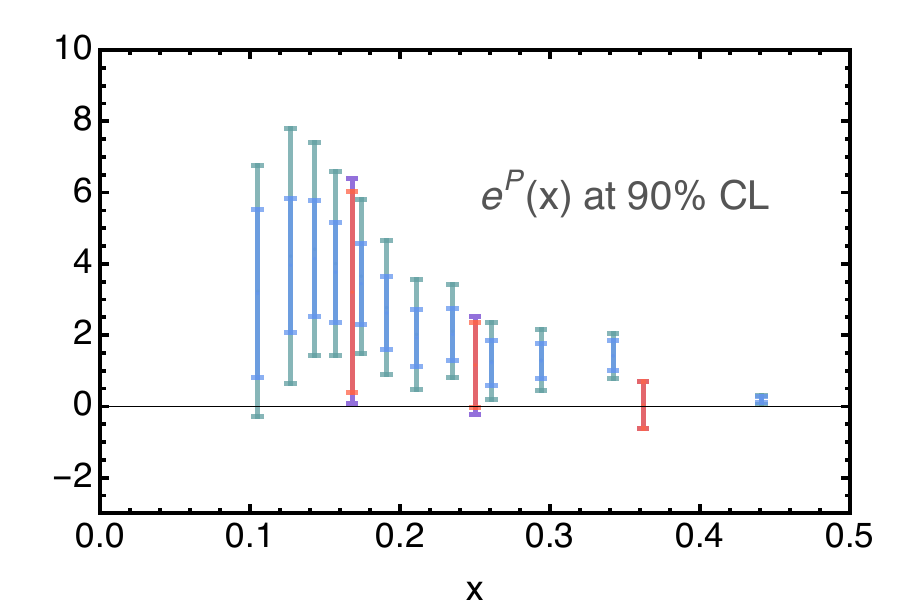}
\caption{The scalar PDF $e(x)$ for the proton combination at 90$\%$ CL. The inner bars represent the contribution from the $0^{th}$ approximation, corrected by the twist-3 contributions for the fragmentation sector for the outer bars. The red-like bars illustrates the extraction from the CLAS data, and the green-hued bars, for the CLAS12 data. }
\label{fig:eP_all_WW+lead}
\end{figure}

The expression for the scalar PDF, in this case, is simply
\begin{eqnarray}
x_i^2\, e^{\rm P}(x_i)&=& -\frac{A(y_i)}{W(y_i)}\frac{Q_i}{M}\, A_{LU}^{\sin \phi_R} \left( x_i ; Q_i^2, y_i\right) \frac{1}{9} \,\frac{ 4 x_i f_1^{u+\bar u} (x_i)\,n_{u, i}+  x_i f_1^{d+\bar d}(x_i)\, n_{d, i}%+  x_i f_1^{s}(x_i)\, n_{s, i}
}{n_{u, i}^{\uparrow}}%\quad.
\label{eq:therealextract}
\end{eqnarray}
The proton combination $e^{\rm P}(x)$ is  depicted by the inner bars in Fig.~\ref{fig:eP_all_WW+lead} at $90\%$ CL.
Notice that the range of integration in $M_h$  goes beyond the range of known validity of the DiFF data set, {\it i.e.} the Belle data with $ 2m_{\pi}<M_h<1.29$ GeV, for which extrapolation from the DiFF fits as been employed for $M_h>1.29$ GeV.
In this scenario, the scalar PDF is clearly non-zero for CLAS12 data, but still marginally compatible with zero as concerns the CLAS data at 6GeV. In this subsection, we have used the MSTW08LO unpolarized PDF set~\cite{Martin:2009iq}.

%%%%%%%%%%%%%%%%%%%%%%%%%%%%%%%%%%%%%%%%%%%%%%%%%%%%%%%%%%%%%%%%%%%%%%%%%%%%%%%%%%%
%%%%%%%%%%%%%%%%%%%%%%%%%%%%%%%%%%%%%%%%%%

%%%%%%
\subsubsection{Beyond the $0^{th}$-approximation scenario}
\label{subsec:beyond}

Meaningful interpretations of the origin of twist-3 effects rely on the inclusion of the quark-gluon-quark components in the equations of motion. As such, the analysis presented as the $0^{th}$ approximation enters in conflict with low-energy theory predictions. 
While our analysis is performed with low-energy  data, the framework of pQCD holds, though we use it at Born level, and the pQCD degrees of freedom will prevail over the nonperturbative ones. In that sense, in this Section, we aim to account for genuine nonperturbative effects from the fragmentation sector. 
On the one hand, this will allow for a more accurate extraction of the $x$-dependence of the scalar PDF ; on the other hand, its interpretation could be extended beyond the Wandzura-Wilczek reduction to twist-$2$ contributions.

In model evaluations, the chiral-even twist-3 interference fragmentation function that contributes in Eqs.~(\ref{eq:alu}, \ref{eq:alu_x_fl}) was shown to be smaller than $H_1^{\sphericalangle}$ and, most importantly, possibly of opposite sign~\cite{Yang:2019aan}. 
The absolute sign of dihadron functions cannot be determined from the specific model calculation, nor can it be from electron-positron experiments~\cite{Courtoy:2012ry}. No independent data set is available yet to explore the behavior of $\tilde{G}^{\sphericalangle}$. 

In App.~\ref{sec:COMPASS}, we examine another twist-3 dihadron fragmentation, $\tilde{D}^{\sphericalangle}$, that is conceivably accessible from COMPASS data. 
The measurement of $\phi_R$ modulations in unpolarized or doubly-polarized collisions\footnote{ See Eq.~(\ref{F_LLcosphi}) below.}  offers a way to address the validity of the $0^{th}$ approximation, and represent an example of how dihadron measurements can be useful for the 
study of quark-gluon correlations in general.
Preliminary data from CLAS indicate that the $\cos\phi_R$ modulation, Eq.~(\ref{F_LLcosphi}), of the Double Spin Asymmetry (DSA)  is very small {\it w.r.t.} the constant term, Eq.~(\ref{F_LL})~\cite{DSA_sergio}. 
This is also confirmed by COMPASS data on dihadron muoproduction off longitudinally-polarized protons~\cite{Sirtl:2016, Sirtl:2017rhi}. Yet COMPASS data lead to a DSA that is much larger than the other twist-3 asymmetries  considered in the present paper. 
For our purpose, COMPASS asymmetries for longitudinally-polarized target will be analyzed in a similar fashion as in Sect.~\ref{subsec:zmpipi} for the reconstruction of the dihadron-related projections for  the  CLAS beam-spin asymmetries. The results are given in App.~\ref{sec:COMPASS}.

Accuracy on twist-3 fragmentation functions would require a dedicated and separate analysis. We do not tackle such a task in this manuscript. Rather, we aim to estimate a proportionality coefficient to the integrated twist-$2$ interference fragmentation function, $n^{\uparrow}$. This is achieved through the estimate of the ratio of integrated DiFFs $n_u^{\uparrow}/n_u$ at the relevant kinematics (provided in Ref.~\cite{Sirtl:2016}) together with the ratio of helicity to unpolarized PDFs, on the $(x,z,M_h)$-triptych\footnote{Notice that COMPASS asymmetries are corrected by the depolarization factor $W(y)/A(y)$, or $K_3$.}.

It has been suggested in the past, {\it i.e.}~\cite{Bacchetta:2003vn}, that combining beam-spin $A_{LU}$ and target-spin $A_{UL}$ asymmetries, both at twist-$3$ level, would facilitate the cancellation of the unknown fragmentation contribution. However, data of dihadron SIDIS off polarized target is not available at this time at CLAS, and vice versa for COMPASS. Moreover, the experimental setup would lead to differences in systematics, for example, as expected from a difference in target polarization.

%%%%%%%%%%%%%%%%%%%%%%%%%%%

In Appendix~\ref{sec:COMPASS}, we argue that an upper value for the twist-3 contribution from the fragmentation sector to the DiFF-projected asymmetry would be given by $\kappa_{M_h}$, whose value can be found in Eq.~(\ref{eq:kappa}). The sign of twist-3 DiFFs being indeterminate for now, we consider that both $+$ and $-$ are possible. 
Then the BSA~(\ref{eq:alu}) becomes
\begin{eqnarray}
A_{LU}^{\sin \phi_R } \left( x_i, M_{h,\, i},  z_i ; Q_i, y_i\right) 
&=&-\frac{W(y_i)}{A(y_i)}\,\frac{M}{Q_i}\,\,\frac{\left[x_i e^P(x_i) \pm\, \kappa_{M_h} f_1^P(x_i)/z_i\right]\, n_{u,\, i}^{\uparrow}   } 
       { \sum_{q=u,d}\, e_q^2\,f_1^q(x_i)\, n_{q, \, i}   }%\quad .
       \nonumber\\
       \label{eq:alu_lead}
\end{eqnarray}
Since the $(z, M_h)$-dependence is integrated,  a non-zero twist-3 PDF becomes manifest in  deviations from the trend in $x$ given by the unpolarized PDF contribution,
\begin{eqnarray}
A_{LU}^{\sin \phi_R }(x_i )\propto \frac{  \left(4\,f_1^{u_V}-  f_1^{d_V}\right)(x_i)}{\left(4\, f_1^{u+\bar{u}}+\,f_1^{d+\bar{d}}\right)(x_i)}\nonumber\\
\label{eq:f1x}
\end{eqnarray}
the trend and size  of which has been estimated with the MSTW08LO set~\cite{Martin:2009iq} as well as from NNPDF2.3 at NLO~\cite{Ball:2012cx} and  CT18NLO~\cite{Hou:2019efy}.\footnote{The use of PDF sets beyond LO is justified in ratios, and allows us to evaluate other sources of uncertainties, while it is not possible for absolute PDFs.} Therefore, going beyond the WW approximation, the BSA  is straightforwardly inverted to get

\begin{eqnarray}
x_i^2\, e^{P}(x_i)&=& -\frac{A(y_i)}{W(y_i)}\frac{Q_i}{M}\, A_{LU}^{\sin \phi_R} \left( x_i,  M_{h,\,i},  z_i ; Q_i^2, y_i\right)\frac{1}{9} \,\frac{ 4 x_i f_1^{u+\bar u} (x_i)\,n_{u, i}+  x_i f_1^{d+\bar d}(x_i)\, n_{d, i}}{n_{u, i}^{\uparrow}}\nonumber\\
&&\mp\kappa_{M_h}\,\frac{x_i}{ z_i}f_1^{P}(x_i)
%\quad,
\label{eq:therealextractBEYOND}
\end{eqnarray}
The results are depicted in Fig.~\ref{fig:eP_all_WW+lead} by the outer bars, which form an envelope around the extraction from the $0^{th}$ approximation due to the indeterminate sign of the twist-3 DiFFs.

The presence of the twist-3 fragmentation contribution, through $\kappa$, slightly modifies the interpretation of the reconstruction of the $(z,M_h)$ dependence in Sect.~\ref{subsec:zmpipi}.

%%%%%%%%%%%%%%%%%%%%%%%%%%%%%%%%%%%%%%%%%%%
\section{Implications for phenomenological analyses}
\label{sec:disc}

We would now like to provide a combined envelope for the final extracted scalar PDF. The results can be cast into three near-Gaussian distributions -- the $N_{\rm rep}$ replicas are almost Gaussianly distributed, there is not minimization involved in our procedure. The first Gaussian stands for the $0^{th}$ approximation, the other two for the next-to-$0^{th}$ approximation with $+$ or $-$ in Eq.~(\ref{eq:alu_lead}).

%%%%%%%%%%%%%%%%%%%%%%%%%%%%%%%%%%%%%%%%%%%%%%%%%%%%%%%%%%%%%%%%%%%%%%%%%%%%%%%%%%%
\begin{figure}[tb]
\includegraphics[scale=.85]{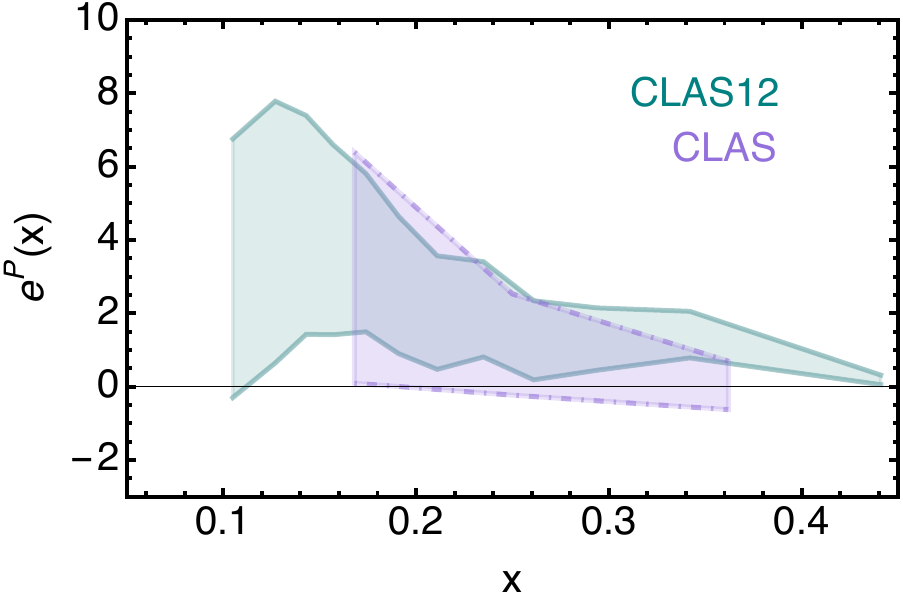}
\includegraphics[scale=.85]{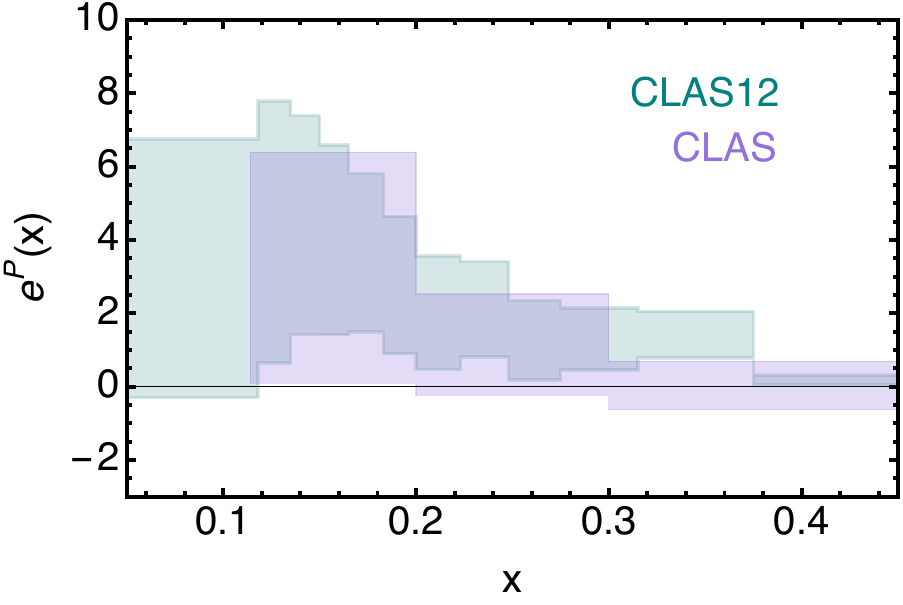}
\caption{Same as in Fig.~\ref{fig:eP_all_WW+lead}, the scalar PDF $e(x)$ for the proton, combined. The final envelopes at 90$\%$ CL are given, for CLAS in purple and CLAS12 in turquoise, in two different formats. On the {\it l.h.s.}, the curves are designed joining the upper and lower values of $e^P(x)$ at the average $x$'s. On the {\it r.h.s.}, the same upper and lower values are shown spanning the full bin interval instead. The results have not been normalized to nor integrated over the bin size. }
\label{fig:exAll}
\end{figure}
%%%%%%%%%%%%%%%%%%%%%%%%%%%%%%%%%%%%%%%%%%
%%%%%%%%%%%%%%%%%%%%%%%%%%%%%%%%%%%%%%%%%%%

The results obtained in the next-to-$0^{th}$ approximation are symmetric around those of the $0^{th}$ approximation. Hence the 90$\%$ CL are easily evaluated either through the usual Monte Carlo prescription (removing $X\%$ of the results for the replicas on both extremes for a $(100-2X)\%$ CL envelope) or processing the sum of the three Gaussians that represent each approximation (selecting the limits that lead to an integral of $X/100$ for an $X\%$ CL envelope). Both approaches agree with one another. No correlations among the three results are considered here, this technicality is beyond the scope of this work. The results are shown in Figs.~\ref{fig:eP_all_WW+lead} and~\ref{fig:exAll} at 90$\%$ CL.

\begin{figure}[tb]
\includegraphics[scale=.85]{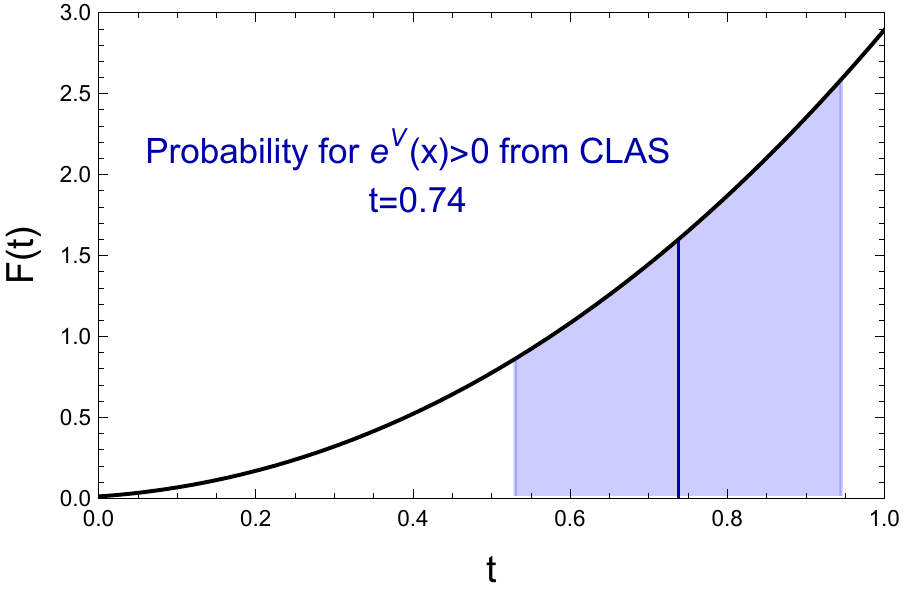}
\includegraphics[scale=.85]{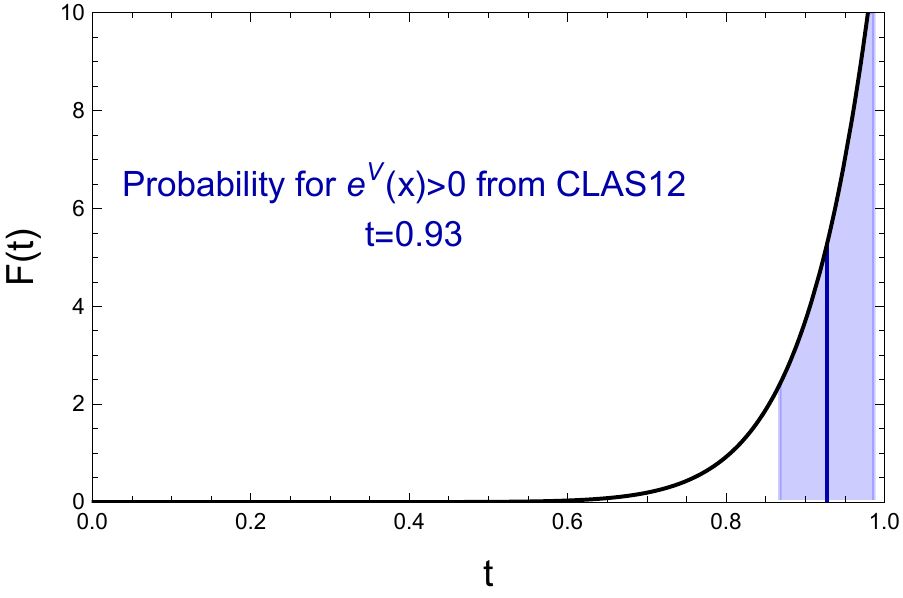}
\caption{Probability for $e^P(x_i)$ to be positive, given as the function $F(t)$ with $t$ an hyperparameter corresponding to the probability that the extracted points  lies above the abscissa. The central value for $t$ is depicted by the thin blue line and quoted in each plot together with the $1\sigma$ uncertainty.  }
\label{fig:Probex}
\end{figure}

We aim to answer an urging question: are twist-3 PDF non-vanishing? More specifically, is the scalar PDF, that obeys no Wandzura-Wilczek reduction, non-zero?  From the CLAS12 data, it is clear that twist-3 observables are non-zero. To answer the question, we estimate the probability for the proton-flavor combined $e(x)$ to be larger than zero, using the reasoning described in Ref.~\cite{Benel:2019mcq}. Through Bayesian statistics, we provide a probability for the extracted PDF to lie inside the positive region of solutions. The results are shown in Fig~\ref{fig:Probex}: CLAS data lead to $e^P(x)>0$ with 74$\%$ probability against 93$\%$ for CLAS12. 

%%%%%%%%%%%%%%%%%%%%%%%%%%%%%%%%%%%%%%%%%%%%%%%%%%%%%%%%%%%%%%%%%%%%%%%%%%%%%%%%%%%
%%%%%%%%%%%%%%%%%%%%%%%%%%%%%%%%%%%%%%%%%%
%%%%%%%%%%%%%%%%%%%%%%%%%%%%%%%%%%%%%%%%%%%
\section{Conclusions}
\label{sec:concl}

We have presented the first extraction of the scalar PDF, a twist-3 object, from dihadron semi-inclusive production using all CLAS data on proton target, employing a minimal set of approximations. The resulting flavor combination is positive at more than 74$\%$ probability. 
The analysis has been carried out through the existing determination of the twist-2 dihadron fragmentation functions~\cite{Radici:2015mwa}, complemented by an extensive yet not thorough analysis of the COMPASS data on similar asymmetries for longitudinally-polarized target~\cite{Sirtl:2016,Sirtl:2017rhi}. Our results improve the preliminary analysis of Ref.~\cite{Courtoy:2014ixa} as well as the  TMD extraction~\cite{Efremov:2002ut}. 
Our methodology differs from the approaches used in the available determinations of $g_T$~\cite{Jimenez-Delgado:2013boa,Sato:2016tuz}, the only collinear twist-3 object accessible in inclusive DIS. As such, our results not only complement but also improve the focus of phenomenological analyses of twist-3 distribution functions.

More data are expected from the CLAS collaboration.   Complementary kinematical regions will be explored at the Electron-Ion Collider (EIC) --mainly small $(x,y)$ regions--
as well as at Jefferson Lab 24 GeV --at slightly larger values of $x$ and $y$ {\it w.r.t} the EIC-- in the future. 
Thanks to the EIC, we will tackle the question of the role of gluon at mid-energies, allowing for a study of evolution of, among other things, twist-3 PDFs. The complexity is twofold: twist-3 contributions are suppressed by $M/Q$ and  the   evolution equations for the scalar PDF, known starting from its second moment~\cite{Balitsky:1996uh,Belitsky:1997zw,Koike:1996bs}, 
needs to be implemented in a practical way. A case for the scalar PDF as the silver channel to access multiparton correlations at the EIC was made in Ref.~\cite{AbdulKhalek:2021gbh}. A similar justification is now being extended to the possibility of a second interaction region at the EIC which would be dedicated to observables sensitive to lower center-of-mass energies~\cite{IR2}. 

With the advent of new experimental data, {\it e.g.}~\cite{CLAS:2021opg}, and facilities, both the collinear and the TMD extractions of $e(x)$ might gain in accuracy. This will be achieved if fragmentation functions, who will play an increasingly important role, are themselves improved. In view of these future developments, we would like to raise the question about the role of the unpolarized PDFs that contribute to the multiplicities (denominators). 
Appreciable conceptual and numerical differences are found for such PDFs extracted more or less stringent cuts in $W$ and $Q^2$~\cite{Accardi:2021ysh}. Twist expansion of observables suggest that  all-purpose global PDFs, extracted from high-energy experiments, should be used. However for such PDFs,  a scale of $Q=1$ GeV belongs to the extrapolation region.
On the other hand, theoretical efforts towards the understanding of TMD factorization of twist-3 will stimulate progress in the TMD sector~\cite{Vladimirov:2021hdn,Ebert:2021jhy}

Complementary to the phenomenological studies, the lattice QCD efforts on evaluating charges and quasi/pseudo-PDFs have been extremely fruitful in the past years, with the premise that  knowledge from both the phenomenological and the lattice side could be bridge for twist-3 PDF in a near future~\cite{Bhattacharya:2020jfj,Braun:2021gvv}.

Our main results, which consist in the $x$ dependence of the scalar PDF depicted in Fig.~\ref{fig:exAll} and of its truncated integral Eq.~(\ref{eq:nx}), will next be used to perform a (global) analysis of the chirally odd twist-3 distribution function $e(x)$~\cite{future},
	\begin{eqnarray}
	e^q(x)&=&\frac{1}{2M}\, \int \frac{d\lambda}{2\pi}\, e^{i\lambda x}\langle P| \bar{\psi}_q(0)\psi_q(\lambda n)|P\rangle,
	\label{eq:ex_def}
	\end{eqnarray}
In that sense, our analysis would pave the way towards a complementary phenomenological way towards specific intrinsic properties of hadrons. The scalar charge is a natural candidate for a long-term goal. It has a dual role as a key information for the understanding of the decomposition of the proton mass as well as
in searches for new physics, along the lines of what was proposed in Ref.~\cite{Courtoy:2015haa} for the tensor charge.

%%%%%%%%%%%%%%%%%%%%%%%%%%%%%%%%%%%%%%%%%%%
\section*{Acknowledgements}

This work has benefited from discussions with D.~Hasch, C.~Lorc\'e, P.~Nadolsky and M.~Wakamatsu.  AC wrote these acknowledgements in memory of M.~V.~Polyakov, whose guidance in the early stage of this work (and, more importantly, the early stage of AC's career) has been decisive.
AC is supported by UNAM Grant No. DGAPA-PAPIIT IN111222 and CONACyT Ciencia de Frontera 2019 No.~51244 (FORDECYT-PRONACES). AM acknowledges financial support from CONACyT through the program Estancias posdoctorales nacionales.

%%%%%%%%%%%%%%%
\appendix

%appendix about DiFF definitions
%%%%%%%%%%%%%%%%%%%%%%%%%%%%%%%%%%%%%%%%%%%
\section{Dihadron fragmentation}
\label{app:diff}
\label{sec:BSA}

%Check all signs and definitions here.\\

At twist-2 level, the $\pi^+\pi^-$-DiFFs relevant  to our study are the unpolarized $D_{1}$ and the chiral-odd $H_{1}^{\open}$.  The $D_1^q$ is the 
unpolarized DiFF describing the hadronization of a parton with flavor $q$ into an unpolarized 
hadron pair plus anything else, averaging over the parton polarization. The $H_1^{\open\, q}$ 
is a chiral- and $T$-odd DiFF describing the correlation between the transverse polarization of the 
fragmenting parton with flavor $q$ and the azimuthal orientation of the plane containing the 
momenta of the detected hadron pair. 
In a Partial Wave Analysis, the physical interpretation of the dominant contribution to $H_1^{\open}$ is related to the interference between relative $p$ and $s$ wave of the pion-pairs, while, for $D_1$, the pion-pairs  are in  relative $s$ waves~\cite{Bacchetta:2002ux}. 

DiFFs depend on the fraction of longitudinal momentum, $z=z_1+z_2$, of fragmenting quark carried by the pion-pair,  on the ratio $\zeta=(z_1-z_2)/z$ ---that can be expressed in terms of the polar angle $\theta$, formed bewteen the direction of the back-to-back emission of the two hadrons  in the center of mass frame and the direction of average momentum of the hadron pair in the target rest frame--- and on the invariant mass of the pair, $M_h$~\cite{Bacchetta:2002ux}.

DiFFs have been studied in models~\cite{Bacchetta:2006un,Casey:2012ux,Matevosyan:2014gea} and have been analyzed  for $\pi^+\pi^-$ production from Belle data~\cite{Courtoy:2012ry}. In particular, $H_1^{\open}$ was extracted from the Artru-Collins asymmetry measured at Belle, using $D_1$ fitted from the output of the Monte Carlo event generator tuned for Belle~\cite{Courtoy:2012ry}.
A functional form at the hadronic scale $Q_0^2=1$~GeV$^2$ was found, fitting the $100$~GeV$^2$ data. The  range of validity of the DiFF fits reflects the kinematic range of the Belle data. In particular, the  integrated range in invariant mass considered for the fit is limited to $2 m_\pi \leq M_h \leq 1.29$ GeV,  the upper cut excluding scarcely populated or frequently empty bins for the Artru-Collins asymmetry. This limit varies bin by bin and the upper limit in $M_h$ can be as low as $0.9$ GeV for $z=0.25$.\footnote{See Fig.~6 of Ref.~\cite{Courtoy:2012ry}.}

To defined dihadron-SIDIS observables, we consider the process
\begin{equation}
  \label{2hsidis}
\ell(l) + N(P) \to \ell(l') + h_1(P_1)+h_2(P_2) + X %,
\end{equation}
where $\ell$ denotes the beam lepton, $N$ the nucleon target,  $h_1$ and $h_2$ the produced hadrons, 
and where four-momenta are given in parentheses. 
We work in the one-photon exchange approximation and neglect the lepton mass.   
The momentum transferred to the nucleon target is $q=l-l'$.
The masses of the  final hadrons are 
$m_1$, $m_2$ and  their momenta  are, respectively, $P_1$, 
$P_2$. The total momentum of the pair is $P_h=P_1+P_2$; the relative momentum $R=(P_1-P_2)/2$ and its component   orthogonal to $P_h$ is  $R_T \equiv R -(R\cdot\hat{P}_h)\hat{P}_h$. 
The  
invariant mass squared of the hadron pair is $P_h^2 = m_{hh}^2$.  The SIDIS process is defined by the kinematic variables:
\begin{eqnarray}
&&x= \frac{Q^2}{2\,P\cdot q}\equiv x_B\quad ,\quad y = \frac{P \cdot q}{P \cdot l}%\quad,
\nonumber\\
&&z = \frac{P \cdot P_h}{P\cdot q}=z_1+z_2%\quad.
\label{eqn:kinem}
\end{eqnarray}
 The kinematics and the definition of the angles can be be found in, {\it e.g.}, Refs.~\cite{Bacchetta:2003vn,Bacchetta:2012ty}. We mention the azimuthal angle $\phi_R$ formed between the leptonic plane and the hadronic plane  identified by the vector $R_T$ and the virtual photon direction. The cross section for two particle SIDIS can be written in terms of modulations in the azimuthal angle $\phi_R$~\cite{Bacchetta:2006tn}.

The Trento convention~\cite{Bacchetta:2004jz} is used in recent publications, as opposed to the original paper on dihadron fragmentation at subleading twist~\cite{Bacchetta:2003vn}. 
%The asymmetries are given by, for the single-spin asymmetry on transversely polarized target that provides the access to the collinear transversity,
%
%\begin{equation}
%A_{\mathrm{SSA}} (x, z, M_h; Q, y) = - \frac{B(y)}{A(y)} \,\frac{|\bf{R} |}{M_h} \, 
%\frac{ \sum_q\, e_q^2\, h_1^q(x; Q^2)\, H_1^{\open\, q}(z, M_h; Q^2)    } 
 %       { \sum_q\, e_q^2\, f_1^q(x; Q^2)\, D_{1}^q (z, M_h; Q^2) } \quad,
%\label{e:ssa}
%\end{equation} 
%
The sign of $n_u^{\uparrow}/n_u$, defined in Eq.~(\ref{e:nq}), was chosen to be negative to ensure a positive transversity PDF from the single-spin asymmtry off transversely polarized target~\cite{Bacchetta:2011ip}.
%for which the overall minus sign is compensated by the choice for . 
This is allowed from the analysis of the Artru-Collins asymmetry of $[n_u^{\uparrow}]^2/[n_u]^2$ from which the sign cannot be determined.
Similarly, the beam-spin asymmetry, Eq.~(\ref{eq:alu}), exhibits an overall minus sign and will result in a positive combination $A_{LU}^{\sin \phi_R}\times n_u/n_u^{\uparrow}$. 

At twist-3, the number of DiFFs increases. In particular there are four {\it genuine}  twist-3 DiFFs, $\widetilde{D}^{\open},\, \widetilde{G}^{\open}, \, \widetilde{E}$ and $\widetilde{H}$~\cite{Bacchetta:2003vn}.  
The functions $\widetilde{D}^{\open},\, \widetilde{G}^{\open}$ are also Interference Fragmentation Functions, like $H_{1}^{\open}$, and are explored in App.~\ref{sec:COMPASS}. The genuine twist-3 DiFFs  describe the fragmentation of a quark, the propagator of which is corrected by gluon fields up to order ${\cal O}(1/Q)$.  They vanish in the Wandzura-Wilzcek approximation.
%
%
%%%%%%%%%%%%%%%%%%%%%%%
%\section{Beam-spin asymmetry in SIDIS off proton target}
%\label{sec:BSA}

The twist-3 DiFFs appear, in observable, starting from the subleading order in $M/Q$, paired with twist-2 objects.
%To define the relevant asymmetries at subleading order, 
%
In the limit $m_{hh}^2 \ll Q^2$ the structure functions of interest can be written in terms of PDFs and DiFFs, to leading-order, in the 
following way~\cite{Bacchetta:2003vn}\footnote{There is minus sign difference in some structure functions w.r.t. Ref.~\cite{Bacchetta:2003vn} due to the Trento conventions published in 2004~\cite{Bacchetta:2004jz}, as mentioned above.}
\begin{eqnarray} 
\nonumber\\
\label{F_UUT}
\hspace{-3mm}
F_{UU ,T} & = &\sum_q e_q^2\;x f_1^q(x)\, D_1^q%\,,
\\
%\label{F_UUcosphi}
%\hspace{-3mm}
%F_{UU}^{\cos\phi_R} & =&- \sum_q e_q^2\;x \frac{|{\bf R}| \sin \theta}{Q}\, \frac{1}{z}\, f_1^q(x)\,
%\widetilde{D}^{\open\, q}\,,\nonumber\\ 
%\\
\label{F_LUsinphi} 
\hspace{-3mm}
F_{LU}^{\sin\phi_R} &  = &-\sum_q e_q^2\;x\frac{|{\bf R}| \sin \theta}{Q}\,
\biggl[
    \frac{M}{m_{hh}}\,x\, e^q(x)\, H_1^{\open\, q}
    +\frac{1}{z}\,f_1^q(x)\,\widetilde{G}^{\open\, q}\biggr]%\,,
\\
\hspace{-3mm}
\label{F_LL}
F_{LL}  & = &\sum_q e_q^2\;x g_1^q(x)\, D_1^q%, \phantom{\biggl[ \biggr]}
\\
\hspace{-3mm}
\label{F_LLcosphi}
F_{LL}^{\cos \phi_R} & =&- \sum_q e_q^2\;x\frac{|\bf R| \sin \theta}{Q}\,
   \frac{1}{z}\,g_1^q(x)\,\widetilde{D}^{\open\, q} %, 
   \phantom{\biggl[ \biggr]}\\
   \nonumber
   \end{eqnarray} 
with the first subindex of the structure function corresponding to the beam polarization, the second to the target. 
All the DiFFs are functions of $\bigl(z,\cos \theta, m_{hh}\bigr)$. When extracting PDFs from the data, a multiplicative factor representing the average value for $\sin \theta$, which lies in the neighborhood of 1, from each experimental set is taken into account. The contribution from higher partial waves has been studied extensively in both CLAS and CLAS12 analyses, concluding that, for the BSA, the first term of the expansion was still dominant.
%
%Up to date, there is no published experimental information about higher-twist DiFFs besides the potential involvement of subleading fragmentation in the beam-spin asymmetry studied in this paper, as discussed in App.~\ref{sec:COMPASS}.
\\

%%%%%%%%%%%%%%%%%
The flavor combination involved in asymmetries is readily worked out, for both the numerator and denominator of the asymmetries, %involve a sum over flavor that can be further developed 
using the following hypotheses~\cite{Courtoy:2012ry},
\begin{itemize}
\item The charm contribution to $f_1^{q=c}(x)$ is negligible {\it w.r.t} $q=u, \,d, \,s$ at JLab scales. We also neglet to strange contribution.
	
\item Invoking charge conjugation  yields to
    \begin{eqnarray}
	D_{1}^{u\to\pi^+ \pi^-} = D_{1}^{\bar{u}\to\pi^+ \pi^-}\,,&&\qquad
	D_{1}^{d\to\pi^+ \pi^-}=D_{1}^{\bar{d}\to\pi^+ \pi^-}%\quad ;
	\label{eq:D1_CC}
	\end{eqnarray}
together with isospin symmetry between $(\pi^+\pi^-)$ and $(\pi^-\pi^+)$	
	\begin{eqnarray}
	H_{1}^{\open u\to\pi^+ \pi^-} &=& - H_{1}^{\open d\to\pi^+ \pi^-} = - H_{1}^{\open \bar{u}\to\pi^+ \pi^-} = H_{1}^{\open \bar{d}\to\pi^+ \pi^-} %\quad ,
	\label{eq:H1_IS}
	\end{eqnarray}
	and similarly for the higher-twist $\tilde{D}^{\open}$ and $\tilde{G}^{\open}$.
	
\item The Interference FF for strange and charm is zero as there is no interference from sea quarks~\cite{Bacchetta:2006un}. 
	For both $\tilde{D}^{\open}$ and $\tilde{G}^{\open}$, we expect the same interpretation for the sea and gluon contributions as for $H_1^{\open}$.
\end{itemize}
%%%%%%%%%%%%%%%%%
%\input{BSA_def}
%appendix about the asymmetries in compass
\section{Analysis of the azimuthal asymmetries on longitudinally polarized protons from COMPASS}
\label{sec:COMPASS}

While estimating the magnitude of the twist-3 DiFFs and, hence, their contribution to the BSA in CLAS and CLAS12, it became important to complement with results from independent observables, whose access could shed light on the size of those contributions. In this Appendix, we analyze the azimuthal asymmetries for dihadron production in SIDIS off polarized proton target, obtained in COMPASS~\cite{Sirtl:2016,Sirtl:2017rhi}. The impact of this analysis is described in the main body of the manuscript, Sec.~\ref{subsec:beyond}.

The asymmetries for longitudinally-polarized targets involve the chiral-odd and unpolarized twist-2 DiFFs, as well as two different twist-3 dihadron fragmentation functions, $\tilde{G}^{\sphericalangle}$ and $\tilde{D}^{\sphericalangle}$. 
The modulations leading to asymmetries on  a longitudinal-target in semi-inclusive dihadron production at twist $3$ read~\cite{Bacchetta:2003vn},
\begin{eqnarray}
A_{UL}^{\sin \phi_R}(x, M_h, z; Q, y)&=&- \frac{W(y)}{A(y)}\,\frac{M}{Q}\,\frac{|\bf{R}|}{M_h}\,
\frac{ \sum_q\, e_q^2\, \left( x h_L^q(x)\, H_1^{\sphericalangle,\, q}(z, M_h)+\frac{M_h}{z M} \,  g_1^q(x)\, 
          \tilde{G}_{sp}^{\sphericalangle, q}(z, M_h)  \right) } 
       { \sum_q\, e_q^2\,f_1^q(x)\, D_{1,ss+pp}^q (z, M_h) }\nonumber\\
       \label{eq:aul}\\
A_{LL}^{\cos \phi_R}(x, M_h, z; Q, y)&=&\frac{W(y)}{A(y)}\,\frac{M}{Q}\,\frac{|\bf{R}|}{M_h}\,
\frac{ \sum_q\, e_q^2\, \left( x e_L^q(x)\, H_1^{\sphericalangle,\, q}(z, M_h)-\frac{M_h}{z M} \,  g_1^q(x)\, 
          \tilde{D}_{sp}^{\sphericalangle, q}(z, M_h)  \right) } 
       { \sum_q\, e_q^2\,f_1^q(x)\, D_{1,ss+pp}^q (z, M_h) }\nonumber\\
       \label{eq:all}
\end{eqnarray}
where we have obviated the dependence in $Q^2$ in the distribution and fragmentation functions. In a working hypothesis considering only $T$-even distribution functions, the double-spin asymmetry in Eq.~(\ref{eq:all}) is reduced to a single term, that containing the twist-3 dihadron fragmentation function.

The total asymmetries quoted in Ref.~\cite{Sirtl:2016, Sirtl:2017rhi} are, respectively, $A_{UL}^{\sin \phi_R}=0.005\pm 0.001$ (to which a $0.001$ of systematic uncertainties is added) and $A_{LL}^{\cos \phi_R}=0.013\pm 0.006$ (to which a $0.005$ of systematic uncertainties is added). 
In absence of proper phenomenological analyses, this result accommodates for various  interpretations. Both asymmetries involve the same twist-2 objects (that are both known phenomenologically up to uncertainties) but differ in the twist-3 description. 
The fact that the single-spin asymmetry is much smaller than the double-spin asymmetry might come from either a cancellation between the respective terms in Eq.~(\ref{eq:aul}) or an enhancement due to large twist-3 effects in the double-spin asymmetry, Eq.~(\ref{eq:all}). 
The latter could either mean that the twist-3 fragmentation function is large or could come from the combined effect of a non-negligible $T$-odd PDF and a twist-3 DiFF. The asymmetry is indeed larger than expected, but so are the uncertainties. Hints of a possible answer can be found through the model evaluations of both twist-3 fragmentation functions in the spectator model, respectively in Ref.~\cite{Yang:2019aan} and Ref.~\cite{Luo:2019frz}. While a cancellation for the single-spin asymmetry is not favored -- except in the low invariant mass region for the $M_h$ projection, the results of Ref.~\cite{Luo:2019frz} imply that the double-spin asymmetry is not well reproduced in that particular model and with the hypothesis in which only $T$-even distribution functions contribute.

In this work, we do not aim to reply to the previously asked question, but rather to  estimate  the order of magnitude of the twist-3 DiFF $\tilde{G}^{\sphericalangle}$ from those data. For that purpose, we will explore the projections of the usual triptych of variables, based on our knowledge of the twist-2 objects involved in Eqs.~(\ref{eq:aul}-\ref{eq:all}). The projections are obtained as described in Sec.~\ref{subsec:zmpipi}. We have used the helicity PDF from {\sc \small NNPDFpol11\_100}~\cite{Nocera:2014gqa}. For consistency, the unpolarized PDF entering the denominator is chosen from the same group (we use NNPDF2.3~\cite{Ball:2012cx}). In a first time, the twist-3 fragmentation function is chosen to be proportional to the chiral-odd twist-2 DiFF, {\it i.e.}
\begin{eqnarray}
n^{\tilde D}&\propto& n^{\tilde G}\propto n^{\uparrow}
%\label{eq:approxIFF}
\end{eqnarray}
However, this rough approximation is not fully supported by model evaluations.
The distinctive behavior of the invariant mass of DiFF becomes a criteria for estimating the order of magnitude of the proportionality factor, $\kappa$. While that behavior could be similar to the twist-2 IFF for $\tilde{D}^{\sphericalangle}$, $\tilde{G}^{\sphericalangle}$ could exhibit a node in $M_h$~\cite{Luo:2019frz,Yang:2019aan}. Both twist-3 DiFFs are of more than an order of magnitude smaller than the twist-2 analog in the spectator model. While it should be noticed that said evaluation does not explicitly account for typical higher-twist degrees of freedom, the estimated shape of the invariant mass can serve as a guide for our approximations. 

We consider the parameter $\kappa$ as an upper bound for twist-3 DiFFs in a specific kinematic region, so that 
\begin{eqnarray}
n^{\tilde D}=\kappa n^{\uparrow}&>&n^{\tilde G}
\label{eq:approxIFF}
\end{eqnarray}

\begin{figure}[tb]
\includegraphics[scale=0.55]{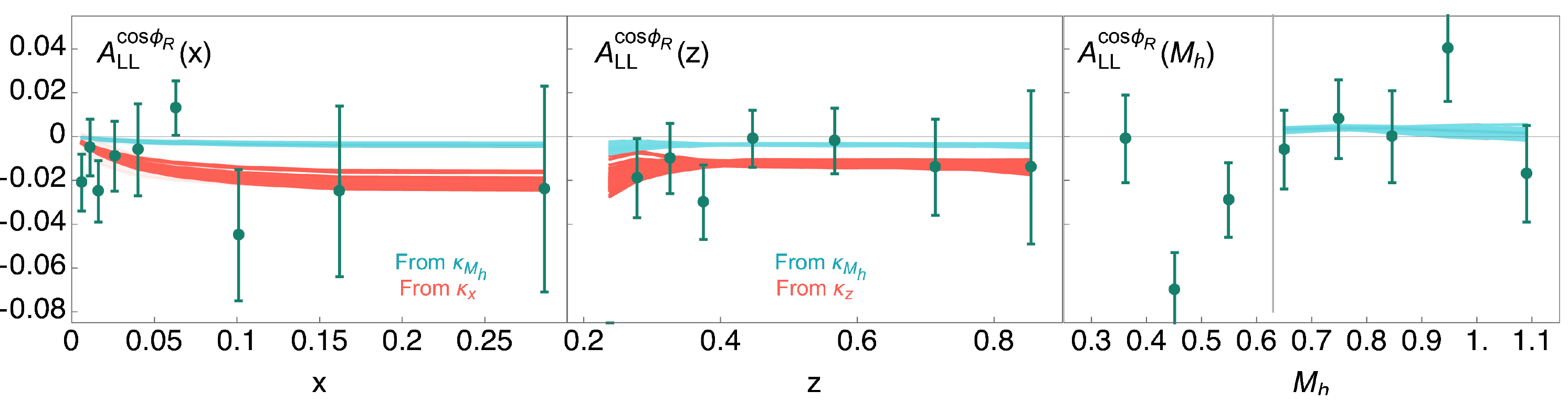}
\caption{The asymmetry $A_{LL}^{\cos\phi_R}$ from COMPASS~\cite{Sirtl:2016}, green points (only statistical uncertainties shown), compared to the reconstructed asymmetry from Eq.~\ref{eq:all} for which we have used that the twist-3 DiFF is proportional to the twist-2 Interference Fragmentation Function through a nornalization factor $\kappa$. Results for $\kappa_{M_h}$ are shown in cyan and for $\kappa_{x/z}$ in red (see text). The light  bands represents to uncertainty coming from the ratio $g_1^{val}/f_1^{\Sigma}$ from {\sc \small NNPDFpol11\_100} and {\sc\small NNPDF23\_nlo}, the  curves represent the bootstrapping results for the DiFFs~\cite{Radici:2015mwa}.}
\label{fig:ALLcosphiR}
\end{figure}

On the right panel of Fig.~\ref{fig:ALLcosphiR}, it can be see that the $M_h$ behavior of the double-spin asymmetry (green points) can clearly not be  reproduced by  twist-2 DiFFs, that is using Eq.~(\ref{eq:approxIFF}). There are hints of contributions from $K^0$ and from the enhanced region of the $\rho$.\footnote{Accounting for the strange DiFF contributions would not improve that comparison so much as to consider it.} Nonetheless, no sign change is predicted from twist-2 DiFFs. We conclude that either the true behavior of $\tilde{D}^{\sphericalangle}$ diverges from that obtained in the spectator model or that the two terms of Eq.~(\ref{eq:all}) interplay. Both observations are supported by a total asymmetry that is large {\it w.r.t.} the twist-3 azimuthal asymmetry, Eq.~(\ref{eq:aul}), reported by COMPASS.

 %%%%%%%%%%%%%%%%%%%%%%%%%%%%%%%%%%%%%%%%%%
%%%%%%%%%%%%%%%%%%%%%%%%%%%%%%%%%%%%%%%%%%

To determine the sought-for factor $\kappa$ from the COMPASS data on longitudinally-polarized target, we will restrict the analysis to reproducing the invariant-mass projection of the double-spin asymmetry for $M_h~>~0.63$ GeV. This prescription was proposed by CLAS12 to focus on the vector meson region~\cite{Hayward:2021psm}. Reconstructing the asymmetry for bins with $M_h~>~0.63$ GeV and comparing with Eqs.~(\ref{eq:all}) for which we have considered only the second term and approximated it by Eq.~(\ref{eq:approxIFF}), we find 
\begin{eqnarray}
\kappa_{M_h}^{\rho}=-0.0818
\label{eq:kappa_good}
\end{eqnarray}
The cyan curves on the right panel of Fig.~\ref{fig:ALLcosphiR} represent the above-described result. The obtained asymmetry is positive in this case.

\begin{figure}[tb]
\includegraphics[scale=0.85]{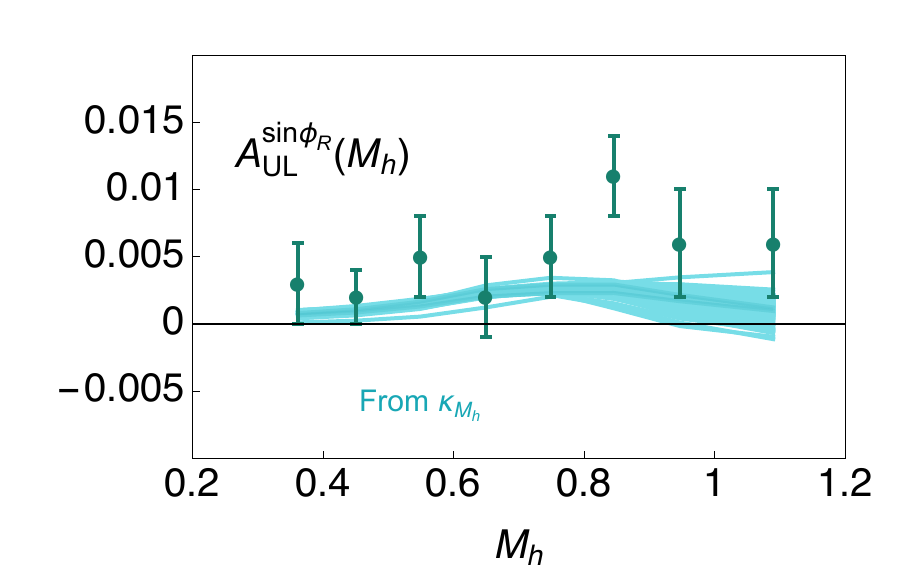}
\caption{The $M_h$-projection of the asymmetry $A_{UL}^{\sin\phi_R}$ from COMPASS~\cite{Sirtl:2016}, green points (only statistical uncertainties shown), compared to the reconstructed asymmetry from Eq.~\ref{eq:all} for which we have used that the twist-3 DiFF is proportional to the twist-2 Interference Fragmentation Function through the nornalization factor  $\kappa_{M_h}$. }
\label{fig:AULsinphiR}
\end{figure}

As mentioned in the previous Section, nor the model explored for DiFFs nor the phenomenological analysis from $e^+e^-$ allow to determine  absolutely the sign of dihadron fragmentation functions.
We can therefore extend the analysis to the lower-$M_h$ region, applying the same technique as before but avoiding the narrow ``kaon" peak. The resulting asymmetry turns negative and
\begin{eqnarray}
\kappa_{M_h}^{\mbox{\tiny all}}=0.0562
\end{eqnarray}
An estimate of the distance between the COMPASS data and the asymmetry reconstructed through Eq.~(\ref{eq:approxIFF}) would reveal a worse outcome for that latter result. Since our goal is to set a conservative estimate on the contribution from twist-3 DiFFs, we adopt the largest absolute $\kappa$,
\begin{eqnarray}
\kappa_{M_h}= \mbox{Max}\left(|\kappa_{M_h}^{\mbox{\tiny all}}|,|\kappa_{M_h}^{\rho}|\right)
\label{eq:kappa}
\end{eqnarray}
The cyan curves on the left and middle panels of Fig.~\ref{fig:ALLcosphiR} represent that choice (with a minus sign). We also show the $M_h$ projection for the single-spin asymmetry in Fig.~\ref{fig:AULsinphiR}: the determined $\kappa_{M_h}$ is in agreement with the reconstruction of $A_{UL}^{\sin\phi_R}(M_h)$ at larger $M_h$ values.

For consistency, we have also aimed to reconstruct the $x$ and $z$ projections of the double-spin asymmetry. The value of Eq.~(\ref{eq:kappa}) reproduces those projections adequately (in cyan on the left and central panels of Fig.~\ref{fig:ALLcosphiR}). However, the iteration of the reconstruction of the integrated asymmetry for those projections leads to larger values of $\kappa_{x,z}$, as depicted in red in Fig.~\ref{fig:ALLcosphiR}. Large error bars and smooth behaviors of the PDFs and the $z$-dependence of the DiFFs do not allow to univocally  validate our approximations. 
All numerical results discussed here should be corrected by uncertainties. However, given the large statistical errors of the data (both from CLAS and COMPASS), it is not essential  to refine our study of $\kappa$. 

 %%%%%%%%%%%%%%%%%%%%%%%%%%%%%%%%%%%%%%%%%%
%%%%%%%%%%%%%%%%%%%%%%%%%%%%%%%%%%%%%%%%%%
%%%%% 

%
The scaling factor $\kappa$ plays a crucial role in our analysis.
 For the $x$-projected asymmetry, that factor multiplied by the known ratio $n_u^{\uparrow}/n_u$ represents all the dependence on the fragmentation part, {\it i.e.} the PDFs have been singled out. While the cyan curves reflect the behavior of $g_1^V/f_1^{\Sigma}$, a contribution from another PDF cannot be excluded.  A similar conclusion can be drawn for the $z$-dependent projection. A effect of phase-space integration, when comparing each panel, will affect the true value of the DiFFs. This statement holds whether we consider twist-2 only or twist-2 and twist-3 combinations. In particular, a reduction of $\kappa$ could be relevant when considering the CLAS data, see Table~\ref{tab:kin}. However, since the kinematic range of COMPASS is the largest and our goal is to provide for an upper limit on the  contribution coming from twist-3 fragmentation, we do not consider such kinematical effects here. 

Finally, we want to emphasize that the results obtained with the COMPASS data  succeeds to the (very) preliminary results obtained from CLAS double-spin asymmetry~\cite{DSA_sergio} -- and that have been originally quoted in the first version of the present manuscript, leading to a rough estimate of $\kappa\sim 0.2$.

%%%%%%%%%%%%%%%%%%%%%%%%%%%%%%%%%%%

\bibliographystyle{apsrev4-1}
\bibliography{alu_bib}

\end{document}